\documentclass[11pt,twoside]{article}

\usepackage{asp2014}
\usepackage{booktabs}

\aspSuppressVolSlug
\resetcounters

\begin{document}

\title{Minimal Re-computation for Exploratory Data Analysis in Astronomy}

\author{Bojan~Nikolic,$^1$ Des~Small,$^2$ and Mark~Kettenis$^2$}
\affil{$^1$Cavendish Lab., University of Cambridge, United Kingdom; \email{b.nikolic@mrao.cam.ac.uk}}
\affil{$^2$Joint Institute for VLBI ERIC, Dwingeloo, The Netherlands}

\paperauthor{Bojan~Nikolic}{b.nikolic@mrao.cam.ac.uk}{0000-0001-7168-2705}{Author1 Institution}{Author1 Department}{City}{State/Province}{Postal Code}{Country}
\paperauthor{Des~Small}{Author2Email@email.edu}{ORCID_Or_Blank}{Author2 Institution}{Author2 Department}{City}{State/Province}{Postal Code}{Country}
\paperauthor{Mark~Kettenis}{Author3Email@email.edu}{ORCID_Or_Blank}{Author3 Institution}{Author3 Department}{City}{State/Province}{Postal Code}{Country}

\begin{abstract}
  We present a technique to automatically minimise the re-computation
  when a data processing program is iteratively changed, or added to,
  as is often the case in exploratory data analysis in radio
  astronomy. A typical example is flagging and calibration of
  demanding or unusual observations where visual inspection suggests
  improvement to the processing strategy. The technique is based on
  memoization and referentially transparent tasks. We describe a
  prototype implementation for the CASA data reduction package.  This
  technique improves the efficiency of data analysis while reducing
  the possibility for user error and improving the reproducibility of
  the final result. 
\end{abstract}

\section{Introduction}

Notwithstanding the notable successes in automating the reduction and
analysis of data from radio telescopes, the traditional
astronomer-driven data reduction is still common. This typically takes
the form of exploratory, iterative, data reduction where visual
inspection of intermediate or final data products is used to adjust,
or add to, the data processing program. Each adjustment is typically
small and impacts only a subset of the processing as a whole; however
in current systems there is no way automatically re-run only this
subset. Instead the user has the choice to either re-run the whole
program, which can take minutes, hours or even days; or to manually do
the sub-selection of the part of the program which needs to be re-run
and risk introducing errors.

The situation can be easily be improved as we show below, by thin
wrappers around existing data reduction software systems and applying
techniques used in other parts of software engineering. We expect the
benefits will be the greatest during telescope commissioning, during
which the automated pipelines are being developed, and to particularly
demanding observations where careful inspection of flagging,
calibration, CLEANing and uv-weighting may be needed.

\section{Objective}

The use case we consider is an astronomer who is repeatedly running a
data reduction processing job with some change to the processing logic
between each run. The primary objective is that, without any
intervention from the astronomer, only the processing steps whose
results could have changed (because either their parameters or their
input data changed) are re-executed between the runs. We assume the
processing logic is captured in a `script' and the above implies
astronomer does not need to edit the script in anyway to select steps
which are to be executed -- instead the script as a whole is submitted
for every execution.

The rationale for this objective is to, at the same time, improve the
efficiency of data analysis (both in term of the computing time and
the time of the astronomers doing the analysis) while reducing the
possibility of errors that is opened by astronomers manually
sub-selecting parts of the script to be run. By only having a single
script whose logic is incrementally improved we expect reproducibility
and understandability will be improved.

Second objective is that the syntax and semantics of the scripts are
as similar as possible to what astronomers are familiar with. In the
field of radio astronomy at least this means Python, or Python-like
syntax and semantics. The rationale for this is to maximise of the
uptake of this software -- very few would be willing to learn a new
paradigm of computing.

Third objective is modularity of the scripts. The rationale is that if
first the first objective is met, modularity will be both easier and
more desirable.

\section{Approach}

\articlefigure{p3-81_fig1.eps}{fig_class}{Classification of
  techniques for minimising computation for all combinations of
  changing program and data scenarios. The primary scenario we
  consider here is in shaded box with bold text -- input data are
  fixed and the program is changing.}

In the scenario we consider the input data are known and unchanging --
they are the raw data recorded by the telescope -- while the data
processing program is evolving between processing runs. This scenario
can be contrasted with one where the program does not change but some
of the input data changes between runs: the classic example of this is
the {\tt make} program which minimises the re-compilation/re-linking
when a subset of source code files changes.  Another scenario 
considered in a previous work \citep{Small2015} is where the input
data are unknown, i.e., the minimal set of re-computation needs to be
determined without reference to a particular input data. The possible
scenarios are classified in Fig~\ref{fig_class}.

The approach we adopt is to have \emph{referentially transparent
  tasks} at the user level and use \emph{memoization}.  A
referentially transparent task is a task whose call can always be
replaced by its return values.  The important implications for
astronomy are that tasks can not have observable side effects other
then their return values and that tasks can not modify in-place any of
their input variables. So for example, a task to calibrate a dataset
must return a new calibrated dataset rather than modifying the dataset
it has been passed. The downside of such tasks is that a processing
job will require more disk storage space; we describe below how we
ameliorate this.

Memoization \citep{michie1968memo, abelson1996structure} is the
technique of tabling the results of task invocations against their
inputs. In astronomy this means tabling the results against both the
input astronomical data and any adjustable processing parameters. The
technique is commonly used in functional programming languages or when
a functional sub-set of a language is used.  A well known example is
Pacrat parsing \citep{Ford:2002:PPS:583852.581483}. The downside of
memoization is the storage space required for keeping the results,
which we limit with an eviction strategy described below. Another
downside in the general case is the high cost of look up for extremely
fine grained tasks but this is unlikely to apply to typical
astronomical processing scripts and their tasks.

The approach we adopt parallels that put forward in other fields in
software engineering. The most direct inspiration was the Ciel
dataflow execution engine \citep{Murray:2011:CUE:1972457.1972470}. A
more generally familiar system is the {\tt ccache} program used to
minimise recompilation when the Makefiles as well as the source code
files are changing in a large compiled-language system. Another
inspiration was the NiX \citep{DolstraNIX} system, which applies the
methodology to building a whole Linux distribution rather than a
single program. From NiX we take the idea of using the filesystem as a
database keyed by the hash of the processing requested from each task.

\section{Design \& Implementation}

In this work we concentrate on the CASA system
\citep{2007ASPC..376..127M}, probably the widest-used software package
in radio astronomy. A similar approach can easily be applied to
ParselTongue \citep{2006ASPC..351..497K} as we showed earlier
\citep{Small2015}. The strategy we adopt is to wrap existing CASA
interface to make it referentially transparent and to replace the use
of user-defined filenames with variables representing files. We then
apply memoization to this new interface, storing the tabling key as
the filename of the value. Values are evicted according the
Least-Recently-Used (LRU) strategy.

Users interact with CASA predominately via ``tasks'', a mostly
functional Python-based interface where input and output data are
always files on disk. CASA tasks however often modify their input
datasets. We make all such tasks referentially transparent by copying
the input dataset before the task is invoked and using this copy as
the output value of the task.

Since some tasks only modify a small part of the input dataset, this
copy approach can be inefficient on traditional filesystems. For this
reason we use OpenZFS (on Linux) which has the Copy-On-Write
architecture (BTRFS also supports this feature). This means that a
copy of a file (i.e., not just a link to) will not cause any copying
of the on-disk blocks until those specific blocks have been changed
for the first time. In the present case this means that when a task
modifies its input dataset, only the blocks it modifies will actually
use disk space.

In normal usage, CASA users specify the files which contain inputs and
outputs of tasks directly by their concrete filenames. Multiple
approaches can be used to apply memoization in this situation. One
approach is to hash the input datafiles and use hashes for tabling;
another is to infer the dataflow \citep[see ][ for a
description]{Small2015}. The approach we adopt is use variables
instead of direct filenames, and to ensure the actual filenames are
the string representations of the hash the task call that was used to
generate their contents. This approach has two advantages: no hashing
of file contents is needed (a potentially expensive operation) as long
as raw input data have stable filenames and are never modified
in-place (the normal situation in radio-astronomy); and, the resulting
code structure is much easier to modularise since the file variables
are lexically scoped in the same way as the other variables. By
contrast, direct use of filenames is equivalent to a single global
namespace for file variables.

Memoized values are checked for eviction before every call into a
task. The trigger is the free disk space on the filesystem used to
store the values, i.e., values will be evicted if the free space is
less than a certain critical value. The criterion used is the
last-access time stamp of the files. This implements the
least-recently-used cache eviction scheme. When a task checks for the
existence of a memoized value, it updates the last-access time even if
the value is not read (i.e., the descendant task has also been
memoized) -- this is done in order to reduce the chance of early
eviction of roots of long, deep dataflows.

Further information and source code is available at
\url{http://www.mrao.cam.ac.uk/~bn204/sw/recipe}.

\acknowledgements

We are pleased to acknowledge the support of EC FP7 RadioNet3/HILADO
project (Grant Agreement no.  283393) and the EC ASTERICS Project
(Grant Agreement no. 653477). We thank A. Madhavapeddy who via Peter
Braam pointed us toward the memoization approach.

\bibliography{p3-81}  

\end{document}